\pdfoutput=1
\documentclass[aip,jap,reprint,groupaddress,noshowpacs]{revtex4-1}

\usepackage[colorlinks=false,bookmarks=true]{hyper ref}

\usepackage{mathrsfs}
\usepackage{epsfig}
\usepackage{amsmath}
\usepackage{subfigure}
\usepackage{amsfonts}

\begin{document}

%
%

\title{A new perspective on renormalization:  the scattering transformation}
%

%
%

\author{Michael E. Glinsky}
\affiliation{School of Physics, University of Western Australia, Crawley, WA, Australia}
\affiliation{CSIRO, Kensington, WA, Australia}


%
%

\begin{abstract}
The scattering transformation developed by \citet{mallat.11} is put into the perspective of field theory.  It is shown to be a simultaneous transformation of the field and the ``time'' parameter explicitly used in the definition of path integrals central to the Lagrangian approach and in the definition of the ``time'' ordered products used in the Hamiltonian (or canonical) approach.  This transformation preserves the form of the S-matrix as ``time'' ordered products.  The transformed ``time'' coordinate is the inverse ``time'' scale.  This is traditionally the UV cutoff or renormalization parameter in standard approaches to renormalization.  The critical calculation will be the determination of the form of the effective action in this transformed coordinate system.  This action will now be expressed as an integral of a Lagrangian density that is a function of the renormalization parameter or transformed ``time''.  Other symmetries of the action can be explicitly built into the transformation.  It will be demonstrated on a simple 1D $\phi^4$ field theory.  This non-perturbative approach has great potential in possibly being used to renormalize quantum gravity and obtaining expressions for the strongly coupled limit of QCD.
\end{abstract}

%
%

\maketitle

%
%

\section{Introduction}

A long standing practical and technical complication of field theory has been the process of regularization and renormalization of the theory.  Constructing regularizations that isolate the singularities yet respect symmetries of the action has been challenging to do, and solutions have been found on an \emph{ad hoc}, case-by-case, basis.  The ability to renormalize the theory without changing the form of the action has been challenging and also only done on an \emph{ad hoc}, case-by-case, basis.  A change of the independent coordinate from the position to the momentum basis has only been of limited utility.  Ken Wilson spoke about a more direct approach in his 1971 seminal paper in Physical Review \citep{wilson.71}.  When trying to block average the physics he longed for a transformation that was \emph{``a quantitative characterization of a complete orthonormal set of minimal wave packets.  For quantitative purposes one would have to take into account tails of the wave packets which extend outside their assigned cells.  It will be assumed here that one can divide phase space into cells of unit volume in any way one pleases and still be able to construct a corresponding set of minimal wave packets.  There is no guarantee that this is actually possible, and no examples of such a set of wave packets will be given here.''}  He then changed course to develop the renormalization group equations.

Recent advancements in the mathematics of signal and image analysis \citep{mallat.11, bruna.mallat.a.11, bruna.mallat.b.11, bruna.mallat.12, bruna.et.al.13} have presented such a direct approach to renormalization in the form of what Mallat calls a scattering operator or transformation of the signal or image.  It turns out that this name is very well chosen.  This operator is directly related to the S-matrix of the field theory.  The transformation is from a field that is a function of coordinates, one of which is a ``time'' that is used to order both the ``time'' ordered products used to define the S-matrix and the related path integrals, to a transformed field that is a function of the renormalization scale or cutoff.  Additional group symmetries of the action are explicitly built into the operator (i.e., transformation) by Mallat.  The hard work goes into the calculation of the transformed action in the transformed coordinates.  This is really just an expression of the physics as a function of the renormalization scale.   This is closely related to to block averaging and the running coupling constants in standard approaches.  In fact, it is found that the action of a physical system (to leading order in $\hbar$) can be written as the classical action (with quantum fluctuation corrections) as a function of the renormalization scale and a transfer matrix (scale dependent mass of the excitations) as a function of the initial and final renormalization scales.  These functions specify the solutions to the renormalization group equations for the coupling constants.  In general, there are an infinite number of solutions, but it can be reduced to just two if the transfer matrix is a simple transfer function of the quotient of the initial and final renormalization scales.  Such physics are called stationary and lead to multi fractal behavior.  If the transfer matrix is a constant, then the physics is self similar.  More importantly, the classical field and the transfer matrix encode the physics of the system and given a state of the field, can be used to identify the physics.

This paper will first present the scattering transform and its relationship to field theory in Sec. \ref{scattering.transform}.  The formalism for the calculation of the transformed (renormalized) action will be shown along with its final simplified form in terms of the classical field and transfer matrix in Sec. \ref{effective.action}.  After a discussion of its relationship to the renormalization group in Sec. \ref{renormalization.group}, and a discussion of how to build in other symmetries of the action in Sec. \ref{group.symmettries}, a simple example of an application to a 1D $\phi^4$ theory will be shown in Sec. \ref{phi.4.app}.  Conclusions which include a discussion of the possible uses of this formalism will be given in Sec. \ref{conclusions}.

\section{Scattering transformation}
\label{scattering.transform}

\begin{figure}
\noindent\includegraphics[width=15pc]{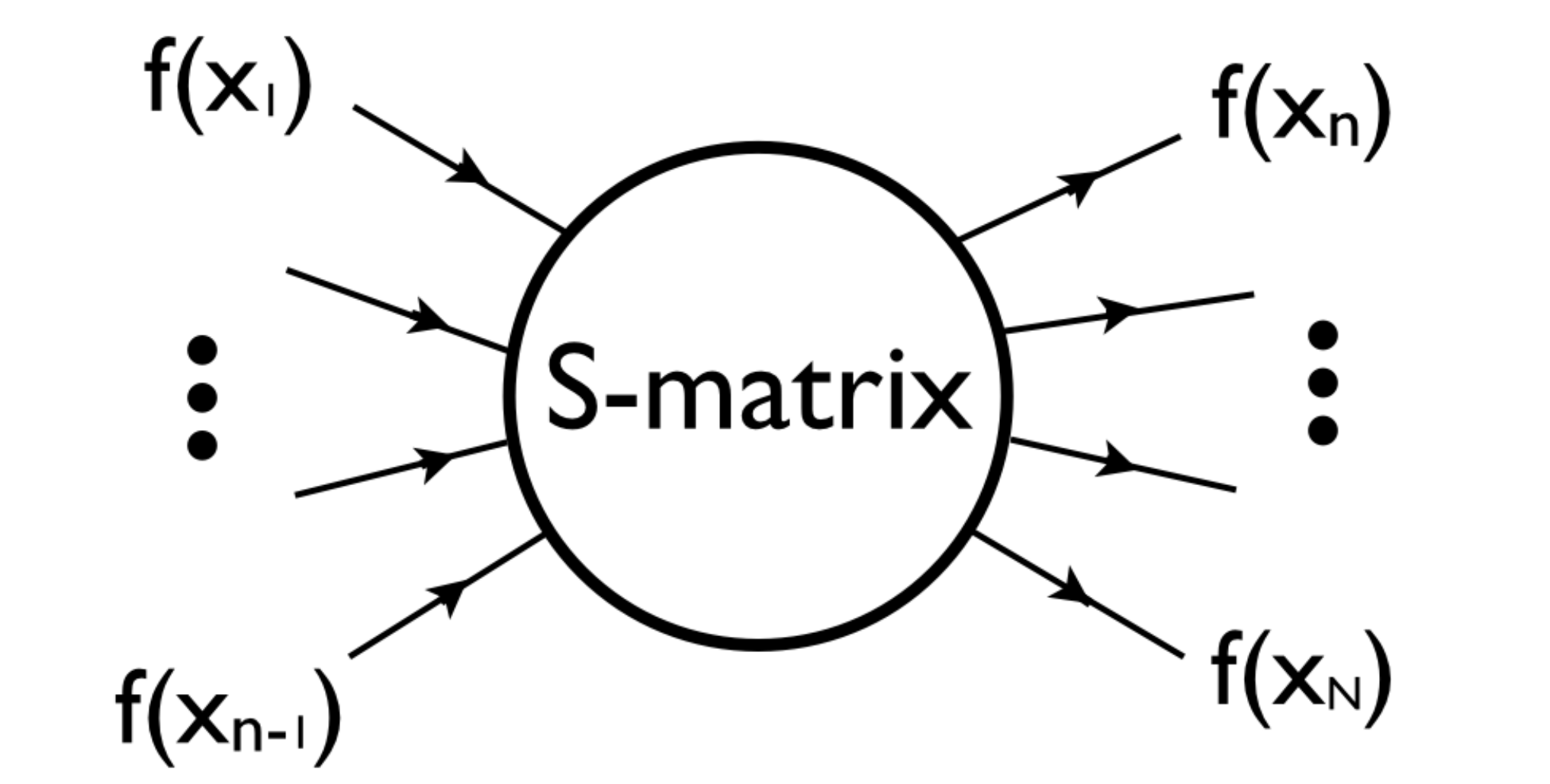}
\caption{\label{fig1} Diagrammatic representation of $N$-particle scattering.}
\end{figure}

Start by considering a field operator, $\hat{f}(x)$, that is a function of a coordinate $x$.\citep{maggiore.10,raymond.90}  A state of the system is $\left|\psi \right>$ and expectation values of operators, $\hat{O}$ are
\begin{equation}
E(\hat{O},\left|\psi \right>)=\left<\hat{O} \right>=\frac{\left<\psi \right| \hat{O} \left|\psi \right>}{\left< \psi | \psi \right>}.
\end{equation}
The S-matrix, $S_m$, is defined as the limit
\begin{equation}
S_m = \lim_{\substack{(x_1,\dots,x_{n-1}) \to -\infty \\ (x_n,\dots,x_N) \to +\infty}}{E(T_x(\hat{f}(x_1) \dots \hat{f}(x_N)),\left|\psi\right>)}
\end{equation}
where $T_x$ is the ``time''  ordered product.  In practice, it is ordered by a chosen coordinate of $x$.  Diagramatically, this is shown in Fig. \ref{fig1}.
This is why the matrix is interpreted as scattering.  In fact, it can be directly related to scattering cross sections.  It can be shown, in this limit, that only the ground state survives so that
\begin{equation}
S_m = \frac{\left<0 \right| T_x(\hat{f}(x_1)\dots \hat{f}(x_N)) \left| 0 \right >}{\left< 0 | 0 \right>},
\end{equation}
where $\left| 0 \right>$ is the ground state.  For a physical system the evolution of $\hat{f}(x)$ can be written as
\begin{equation}
\hat{f}(x_{n+1}) = e^{(i/\hbar) \hat{H} (x_{n+1}-x_n)} \hat{f}(x_n) e^{-(i/\hbar) \hat{H} (x_{n+1}-x_n)},
\end{equation}
or equivalently 
\begin{equation}
\left| f(x_{n+1}) \right> = e^{-(i/\hbar) \hat{H} (x_{n+1}-x_n)} \left| f(x_n) \right>.
\end{equation}
This composition property allows the S-matrix to be written in a Lagrangian form as
\begin{equation}
S_m =\left.  { \frac{1}{Z[J]} \frac{\delta}{\delta J(x_1)} \cdots  \frac{\delta}{\delta J(x_N)} Z[J]} \right|_{J=0}
\end{equation}
where
\begin{equation}
Z[J] = N \int [df(x)] \; e^{(i/\hbar) S_0[f(x)] + (i/\hbar) \int dx J(x) f(x)}
\end{equation}
is the generating function, $J(x)$ are the excitation currents or forces, and
\begin{equation}
S_0[f(x)] = \int dx \; L(f,\partial_x f)
\end{equation}
is the action.  The Lagrangian density can be related to $H$ through the Legendre transform
\begin{equation}
H(\pi,f) = \pi \; \partial_x f - L(f,\partial_x f)
\end{equation}
and
\begin{equation}
\pi \equiv \frac{\partial L}{\partial(\partial_x f)}.
\end{equation}

Now make a joint transformation from $\hat{f}(x) \to \hat{\psi}(\lambda)$ and $\hat{x} \to \hat{\lambda}$ defined by
\begin{equation}
\begin{split}
\left<\psi,\lambda | f,x \right> &= \psi_{\lambda}(x) \star f(x) \\ &= \left| \psi_{\lambda} \star f \right| \; e^{i\varphi(\psi,\lambda,f,x)},
\end{split}
\end{equation}
where $\star$ is the normal convolution operator defined as
\begin{equation}
f \star g = \int dx^\prime f(x^\prime-x) \: g(x^\prime).
\end{equation}
This is distinctly different from a $\hat{x} \to \hat{k}$ change of basis defined by $\left<x | k \right> = e^{ikx}$ (Fourier transform of coordinate) or even a change basis of the field $\hat{f} \to \hat{\pi}$ defined by $\left< \pi | f \right>=e^{i\pi f}$ (a Fourier transform of the field).  Here $\psi_{\lambda}(x)$ are a family of wavelets that can be written as 
\begin{equation}
\psi_{\lambda}(x) \equiv 2^j \; \psi_0 (2^j x)
\end{equation}
where $\lambda_j \equiv 2^j$ and $j>-J$.  For finite $j$, an additional final convolution with a window function,
\begin{equation}
\phi_J (x) \equiv 2^{-J} \; \phi(2^{-J}x)
\end{equation}
will need to be done in order to obtain convergence.  The $J \to \infty$ limit will be taken.  More will be said about this later.  The wavelet, $\psi_0$, satisfies the Littlewood-Pauly condition so that $\psi_{\lambda}(x) \star$ is unitary.  The Mallat scattering operator can now be defined as
\begin{equation}
U[p] \equiv \lim_{J \to \infty} T_{\lambda} (\hat{\psi}(\lambda_1) \dots \hat{\psi}(\lambda_N) \hat{\phi}_J )
\end{equation}
where $p \equiv (\lambda_1,\dots, \lambda_N)$.  One should note that $\varphi(\psi,\lambda,f,x)$ is set equal to zero.  The reason for this will be explained later.  The operator, $U[p]$, operates on a distribution of functions, $F(f(x))$.  The expected value of the scattering operator is given by $E(U[p] \: F(f))$.  It is shown by Mallat that $U[p]$ preserves the norm.  Another way of viewing $U[p]$ is by the result of it acting on a realization of the distribution, $f(x)$, given by
\begin{equation}
U[p] \: f(x) = \lim_{J \to \infty} \left| \left| \left| \left| f \star \psi_{\lambda_1} \right| \star \psi_{\lambda_2} \right| \dots \right| \star \psi_{\lambda_N} \right| \star \phi_J.
\end{equation}

\begin{figure}
\noindent\includegraphics[width=15pc]{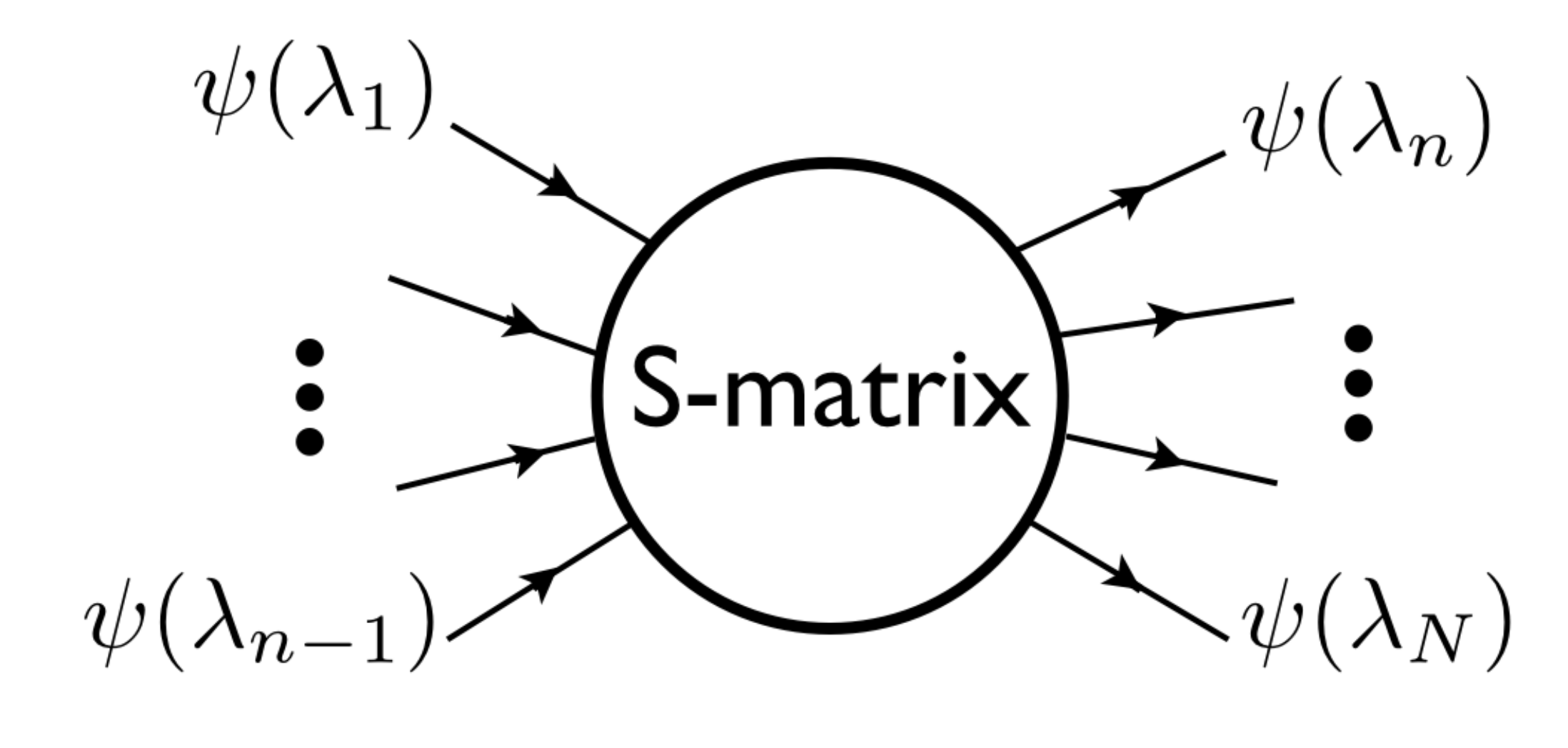}
\caption{\label{fig3} Diagrammatic representation of $N$-particle scattering in the transformed basis.}
\end{figure}

It is apparent that the following correspondence can be made of this scattering operator to field theory.  First, the distribution, $F(f)$, is just the state $\left| \psi \right>$.  In other words, the state is just a distribution of fields.  The expected value of the scattering operator is
\begin{equation}
\begin{split}
E(U[p] \: F(f)) &= E(T_\lambda (\hat{\psi}(\lambda_1) \dots \hat{\psi}(\lambda_N)) \; F(f)) \\
&= E(T_x (\hat{f}(x_1) \dots \hat{f}(x_N)), \left| \psi \right>) \\
&= \frac{\left< 0 \right| T_x (\hat{f}(x_1) \dots \hat{f}(x_N)) \left| 0 \right>}{\left< 0 | 0 \right>} \\
&= S_m,
\end{split}
\end{equation}
that is the S-matrix, shown diagramatically in Fig. \ref{fig3}.
Because of the structure of $p$, it can be mapped onto $\mathbb{R}$ with the following transformation $\lambda = \sum \lambda_i \in \mathbb{R}$.  One can define a generating function
\begin{equation}
\label{psi.gen.function.eqn}
Z[J] = N \int [d\psi (\lambda)] \; e^{(i/\hbar) S_0[\psi (\lambda)] + (i/\hbar) \int dx J(\lambda) \psi (\lambda)}.
\end{equation}
The expectation value of the scattering operator, the S-matrix, can now be calculated using the generating function as
\begin{multline}
E(T_\lambda (\hat{\psi}(\lambda_1)  \dots \hat{\psi}(\lambda_N)) \; F(f)) = \\
\left.  { \frac{1}{Z[J]} \frac{\delta}{\delta J(\lambda_1)}  \cdots  \frac{\delta}{\delta J(\lambda_N)} Z[J]} \right|_{J=0}
\end{multline}
The work is now in deriving the renormalized action, $S_0[\psi(\lambda)]$, given an action, $S_0[f(x)]$.  This will be done in the next section (Sec. \ref{effective.action}) and an example given in Sec. \ref{phi.4.app}.

Two technical points about the action should be noted.  The first is with respect to the factor $\phi_J$ used in the definition of $U[p]$.  In order to get convergence of the path integrals, a Gaussian factor needs to be added.  This is the familiar $i\varepsilon$ prescription that adds a Gaussian convergence term to the oscillating integrals.  Once these integrals are evaluated at finite $\varepsilon$, the $\varepsilon \to 0$ limit is taken.  Therefore, there is a direct correspondence between $i\varepsilon$ and $1/J$.  

Second, it will be found that one will be presented with exponential integrals in the evaluation of the effective renormalized action.  These integrals will be done using the principal of stationary phase which gives an asymptotic expansion for these integrals.  The result will be that the phase of the integrand, $\varphi(\psi,\lambda,f,x)$, must be set to a constant so that the phase will be stationary.  This constant will be chosen to be zero for simplicity.  This is a subtle, but important point.  The modulus in the definition of $U[p]$ does not come from the definition of the transformation, but rather from a condition that is required to evaluate the effective action.  In other words, $U[p]$ can be evaluated for other values of $\varphi$, but the contribution to the expected value will be negligible.  More importantly, this freedom in setting the phase to an arbitrary constant value suggests the existence a gauge boson associated with exchange of this phase.

\section{Calculation of effective action}
\label{effective.action}

Now move onto the evaluation of the effective renormalized action given $S_0[f(x)]$.  One needs to first define what is meant by an effective action.\citep{raymond.90}  Given a generating function, $Z[J]$, the effective action is defined via the Legendre transformation
\begin{equation}
\label{def.effective.action.eqn}
S[\varphi(\lambda)] = - \ln Z[J] + \int d\lambda \; J(\lambda) \; \varphi(\lambda),
\end{equation}
where
\begin{equation}
\varphi(\lambda) \equiv \frac{\delta(\ln {Z[J]} )}{\delta J(\lambda)} = \frac{1}{Z[J]}  \frac{\delta Z[J]}{\delta J(\lambda)},
\end{equation}
\begin{equation}
J(\lambda) = \frac{\delta S[\varphi(\lambda)]}{\delta \varphi(\lambda)},
\end{equation}
and
\begin{equation}
\frac{\delta^2 (\ln Z[J])}{\delta J(\lambda_1) \; \delta J(\lambda_2)} = \frac{\delta^2 S[\varphi(\lambda)]}{\delta \varphi(\lambda_1) \; \delta \varphi(\lambda_2)}.
\end{equation}
This action includes the effects of the quantum fluctuations along the path.  Separate the effective action into two parts, the original action, $S_0$, and the quantum fluctuation part, $S_1$, of order $\hbar$ so that
\begin{equation}
S[\varphi(\lambda)] = S_0[\varphi(\lambda)] + S_1[\varphi(\lambda)].
\end{equation}
We also expand $\varphi(\lambda)$ about $\psi(\lambda)$ so that
\begin{equation}
\xi(\lambda) \equiv \psi(\lambda) - \varphi(\lambda) \ll \varphi(\lambda).
\end{equation}
To second order in $\xi(\lambda)$ one obtains the following equation for $S_1[\varphi(\lambda)]$
\begin{multline}
\exp{(S_1[\varphi(\lambda)])} = \int [d\xi(\lambda)] \; \exp \left( \int d\lambda \; \xi(\lambda) \frac{\delta S_1}{\delta \varphi(\lambda)} \right. \\
\left.  - \frac{1}{2} \int d\lambda \; d\lambda^\prime \; \xi(\lambda) \frac{\delta^2 S_0}{\delta \varphi(\lambda) \delta \varphi(\lambda^\prime)} \xi(\lambda^\prime) \right).
\end{multline}
This is a standard Gaussian integral that can be done but the contour must be deformed so that the phase of the integrand is constant.  This is equivalent to setting the phase, $\varphi(\psi,\lambda,f,x)$, equal to zero.  One gets the following equation for $S_1$
\begin{equation}
2 \: S_1 = \Delta - \int d\lambda \; d\lambda^\prime \; \frac{\delta S_1}{\delta \varphi(\lambda)} \; \gamma(\lambda,\lambda^\prime) \; \frac{\delta S_1}{\delta \varphi(\lambda^\prime)},
\end{equation}
where
\begin{equation}
\gamma(\lambda,\lambda^\prime) \equiv \frac{\delta^2 S_0}{\delta \varphi(\lambda) \delta \varphi(\lambda^\prime)}
\end{equation}
and
\begin{equation}
\Delta \equiv \ln \left| \frac{\gamma}{2 \pi} \right|.
\end{equation}
This can be solved giving $S[\varphi(\lambda)]$.  The mean path, $\varphi_0 (\lambda)$, is given by
\begin{equation}
\left. \frac{\delta S}{\delta \varphi(\lambda)} \right|_{\varphi_0 (\lambda)} = 0,
\end{equation}
leading to the expansion about $\varphi_0 (\lambda)$ to second order
\begin{multline}
S_0[\varphi(\lambda)] = S_0[\varphi_0 (\lambda)] + \frac{1}{2} \int d\lambda \; d\lambda^\prime \left( \varphi(\lambda) - \varphi_0(\lambda) \right) \\
\frac{\delta^2 S_0[\varphi_0 (\lambda)]}{\delta \varphi(\lambda) \delta \varphi(\lambda^\prime)} \left( \varphi(\lambda^\prime) - \varphi_0(\lambda^\prime) \right),
\end{multline}
where
\begin{equation}
S_2(\lambda,\lambda^\prime) \equiv \frac{\delta^2 S_0[\varphi_0 (\lambda)]}{\delta \varphi(\lambda) \delta \varphi(\lambda^\prime)}.
\end{equation}
Now use this expression for the effective action and substitute it into Eq. \eqref{psi.gen.function.eqn} to get
\begin{multline}
Z[J] = N \int [d\varphi(\lambda)] \; \exp \left( \frac{i}{\hbar}S_0[\varphi_0 (\lambda)] + \frac{i}{2 \hbar} \right. \\
\left. \int d\lambda \; d\lambda^\prime \left( \varphi(\lambda) - \varphi_0(\lambda) \right) 
S_2(\lambda,\lambda^\prime) \left( \varphi(\lambda^\prime) - \varphi_0(\lambda^\prime) \right) \right. \\
\left. + \frac{i}{\hbar} \int d\lambda \; J(\lambda) \; \varphi(\lambda) \right).
\end{multline}
By completing the square, it can be written
\begin{multline}
Z[J] = N \exp \left( \frac{i}{\hbar} S_0[\varphi_0 (\lambda)]  
+ \frac{i}{\hbar} \int d\lambda \; J(\lambda) \; \varphi_0 (\lambda) \right. \\
\left. - \frac{i}{2 \hbar} \int d\lambda \; d\lambda^\prime \; J(\lambda) \; S_2^{-1}(\lambda,\lambda^\prime) \; J(\lambda^\prime) \right) \\
\int [d\varphi(\lambda)] \; \exp \left( \frac{i}{2 \hbar} \int d\lambda \; d\lambda^\prime \; \varphi(\lambda) \; S_2(\lambda,\lambda^\prime) \; \varphi(\lambda^\prime) \right).
\end{multline}
Absorbing all terms that are not a function of $J(\lambda)$ into $\tilde{N}$ one obtains
\begin{equation}
Z[J] = \tilde{N} e^{\frac{i}{\hbar} \int d\lambda \; J(\lambda) \; \varphi_0 (\lambda)
- \frac{i}{2 \hbar} \int d\lambda \; d\lambda^\prime \; J(\lambda) \; S_2^{-1}(\lambda,\lambda^\prime) \; J(\lambda^\prime)}.
\end{equation}
The form of $Z[J]$ is determined by specifying $\varphi_0 (\lambda)$ and $S_2^{-1} (\lambda, \lambda^\prime)$.  They can be identified as
\begin{equation}
E(\hat{\psi} (\lambda) F(f)) = \left. \frac{1}{Z[J]} \frac{\delta Z[J]}{\delta J(\lambda)} \right|_{J=0} = \varphi_0 (\lambda)
\end{equation}
and
\begin{equation}
\begin{split}
E(\hat{\psi} (\lambda) \hat{\psi} (\lambda^\prime) F(f)) &= \left. \frac{1}{Z[J]} \frac{\delta^2 Z[J]}{\delta J(\lambda)\delta J(\lambda^\prime)} \right|_{J=0} \\
&= -S_2^{-1} (\lambda,\lambda^\prime).
\end{split}
\end{equation}

To understand better what $S_0[\varphi(\lambda)]$ really is, examine the successive transformation of $f(x)$ by $\hat{\psi} (\lambda_i)$ giving
\begin{multline}
\psi(\lambda) = \int dx_1 \dots dx_{n+1} \; \psi_{\lambda_1}(x_2-x_1) \dots \psi_{\lambda_n}(x_{n+1}-x_n) \: \\
 \phi_J(x_{n+1}) \: f(x_1)
\end{multline}
where $\lambda=(\lambda_1,\dots,\lambda_n)$.

To understand better what is going on, assume that the $\psi_\lambda (x)$ are orthogonal, that is
\begin{equation*}
\int dx^{\prime \prime} \psi_\lambda (x^{\prime \prime}-x) \; \psi_{\lambda^\prime} (x^{\prime \prime}-x^\prime) = \delta(x-x^\prime) \; \delta(\lambda-\lambda^\prime).
\end{equation*}
The functional derivative of $S_0[\psi(\lambda)]$ can be identified as
\begin{equation}
\begin{split}
\frac{\delta S_0[\psi(\lambda)]}{\delta \psi(\lambda)} &= \left< \frac{\partial L}{\partial f} \right>_\lambda +\left< \frac{d}{dx} \frac{\partial L}{\partial(\partial_x f)} \right>_\lambda \\
&= \left< F \right>_\lambda + \left< \dot{\Pi} \right>_\lambda
\end{split}
\end{equation}
where
\begin{multline}
 \left< F \right>_\lambda \equiv \left< \frac{\partial L}{\partial f} \right>_\lambda \equiv \\
 \int dx_1 \dots dx_{n+1} \; \psi_{\lambda_1}(x_2-x_1) \dots \psi_{\lambda_n}(x_{n+1}-x_n) \: \\
  \phi_J(x_{n+1}) \: \frac{\partial L}{\partial f}(x_1)
\end{multline}
and
\begin{multline}
  \left< \dot{\Pi} \right>_\lambda \equiv \left< \frac{d}{dx} \frac{\partial L}{\partial(\partial_x f)} \right>_\lambda \equiv \\
 \int dx_1 \dots dx_{n+1} \; \psi_{\lambda_1}(x_2-x_1) \dots \psi_{\lambda_n}(x_{n+1}-x_n) \: \\
  \phi_J(x_{n+1}) \: \frac{d}{dx} \frac{\partial L}{\partial(\partial_x f)}(x_1).
\end{multline}
This is just a mean field approximation, where there has been a successive averaging at scales $\{ \lambda_1,\dots,\lambda_n \}$.  The first term is the mean generalized force.  The second term is the mean mass times acceleration or change in the generalized momentum.  Now the important second functional derivative is
\begin{equation}
S_2(\lambda,\lambda^\prime) = \frac{\delta^2 S_0[\psi(\lambda)]}{\delta \psi(\lambda) \; \delta \psi(\lambda^\prime)} = \frac{\delta  \left< F \right>_\lambda}{\delta \psi(\lambda^\prime)} + \frac{\delta  \left< \dot{\Pi} \right>_\lambda}{\delta \psi(\lambda^\prime)}.
\end{equation}
Remember that a modulus will need to be inserted after each integration $\int dx_i$ during the successive averaging to keep the dominant contribution to the oscillating path integral.

\section{Relation to the renormalization group}
\label{renormalization.group}

Stand back, take a look a what has been done and compare it to the conventional approach to renormalization.\citep{maggiore.10}  A transformation has been defined that successively averages the physics at different scales, $\{\lambda_1,\dots,\lambda_n\}$.  The scattering matrix is well defined in this new basis and can be calculated from a generating function, $Z[J]$, characterized by two functions, $\varphi_0(\lambda)$ and $S_2^{-1}(\lambda,\lambda^\prime)$.  These are directly related to the action, $S_0[\psi(\lambda)]$, in this averaged basis.  In this basis, everything is a function of $\lambda$, with is a n-tuple, $(\lambda_1,\dots,\lambda_n)$, which can be mapped one-to-one onto $\mathbb{R}$ with the presciption $\lambda=\sum\lambda_i$.  There are no singularities in the calculation of the path integrals, therefore there is no need to renormalize these integrals.  One does have explicit expressions for how the physics changes as a function of inverse scale, $\lambda$.    This is calculated in a well prescribed way for any action, $S_0[f(x)]$.  

This is in contrast to the more conventional approach where one is initially presented with a problem of singularities in the calculation.  The singularities need to be isolated in a way that preserves the symmetries of the system and constants of the physics -- a process called regularization.  This is not an easy task and must be done in an \emph{ad hoc} way on a case-by-case basis.  One then needs to renormalize the physics by dividing out the isolated singularities, again without changing the form of the physics (i.e., action) and preserving the constants of the physics.  This is another difficult task that is done in an \emph{ad hoc} way on a case-by-case basis.  Since there is no systematic way of doing either the regularization or renormalization that is guaranteed to work, a large portion of effort is expended in doing this and many roadblocks are encountered.  One always finds in this process that one parameter is introduced, with units of inverse length in natural units.  These are things like a UV momentum limit, $\Lambda$, in cutoff regularization and the mass, $\mu$, in dimensional regularization.  When the singularities are removed, via renormalization, one finds that the mass, coupling constants, and fields all become functions of the renormalization parameter.  This forms a one parameter group of symmetries generating by changing the scale of the parameter.  This group is normally specified by a set of ODEs with this parameter as the independent variable.  These ODEs are derived from the physics and the fact that the un-renormalized and physical mass, coupling constants, and fields are not functions of the renormalization parameter.  One has gone from renormalizing the calculation (that is, removing singularities), to studying how the physics changes as a function of the renormalization parameter (or scale).  Hence, there is a dual meaning for the word renormalization.  Properly, it is dividing out or renormalizing to remove a singularity.  Practically, it is understanding how the physics changes as a function of the scale of the renormalization.

This now leads to the important conclusion -- what is primary is how the physics changes as a function of scale.  This is the natural basis for the S-matrix.  The singularities in the calculation of the S-matrix in the conventional approach are introduced because of a poor choice of basis.  With the correct choice of basis, which expresses the physics based on scale, there are no singularities or need to renormalize.  This is the new perspective on renormalization -- it is not needed.  The part of the conventional process which survives is the understanding of how the physics ``renormalizes'' as a function of inverse scale, $\lambda$.

Now the obvious question is  ``where are the solutions to the renormalization group equations in this new perspective?''  They are contained in the functions $\varphi_0(\lambda)$ and $S_2^{-1}(\lambda,\lambda^\prime)$.  For instance, assume that there are three renormalization group equations generated by a theory based on a scalar field, a mass, and a coupling constant.  This means that three functions, $a_1(\lambda)$, $a_2(\lambda)$ and $a_3(\lambda)$, specify the scaling behavior of the system.  One can now identify 
\begin{equation}
\varphi_0(\lambda) = a_1(g_1(\lambda))
\end{equation}
and
\begin{equation}
S_2^{-1}(\lambda,\lambda^\prime) = a_2(g_2(\lambda,\lambda^\prime)) + a_3(g_3(\lambda,\lambda^\prime))
\end{equation}
for a proper choice of $g_1(\lambda)$, $g_2(\lambda,\lambda^\prime)$, and $g_3(\lambda,\lambda^\prime)$.  Since physical theories are specified by a small, finite number of fields, masses, and coupling constants; an analysis of the dimensionality of $S_2^{-1}(\lambda,\lambda^\prime)$ will uncover how many there are.  For the case that $S_2^{-1}(\lambda,\lambda^\prime)$ can be written as $S_2^{-1}(\lambda / \lambda^\prime)$, there would be just one scalar field and one mass in the system.  The state of the system would display a multi fractal behavior and the physics would be called stationary.

\section{Building in additional symmetries}
\label{group.symmettries}

Up to this point the scattering transform has respected the renormalization symmetries of the system generated by a change in scale of the classic renormalization.  In fact, $\lambda$ is the parameter of the renormalization group.  This group structure is induced by the composition property of dynamical systems and the resulting exponential form for the path integrals involving the action functional in the exponential, that is $\int [df(x)]e^{(i/\hbar)S_0[f(x)]}$.  The group parameter is associated with the coordinate used to order the physical dynamics, in many cases the time.

A significant difficulty with classic regularization and renormalization is ensuring that the process respects all the symmetries of the physics (that is, the action).  These could be symmetries of the space, field, or combined symmetries.  They could be local (gauge) or global symmetries.  Mallat show how to build additional group symmetries into the transformation so that the renormalization will not break those group symmetries.  This is a straight forward prescription that can be done for any combination of group symmetries.  The result is the introduction of another path variable for each group symmetry parameter and the associated scattering along that path.  This enlarges the scattering space from being indexed by $\lambda=(\lambda_1,\dots,\lambda_n)$ to $\lambda \otimes \tilde{\lambda}$, where $\tilde{\lambda}=(\tilde{\lambda}_1,\dots,\tilde{\lambda}_n)$ is the additional path variable.

\section{Example of 1D $\phi^4$ field theory}
\label{phi.4.app}

Here is a simple example that demonstrates how this renormalization works.  Examine the 1D system with an action that has the basic characteristics of a $\phi^4$ theory
\begin{equation}
S_0[f(x)] = \int dx \; \frac{1}{2} \left( \frac{df}{dx} \right)^2 - \frac{m^2}{2} f^2 - \frac{\beta}{4!} f^4.
\end{equation}
This action when transformed gives
\begin{multline}
\label{phi.4.action.eqn}
S_0[\psi(\lambda)] = \int d\lambda \frac{\lambda^2 - m^2}{2} \psi^2(\lambda) - \frac{\beta}{4!} \int d\lambda \: d\lambda^\prime \: d\lambda^{\prime \prime} \: d\lambda^{\prime \prime \prime} \; \\
A(\lambda,\lambda^\prime,\lambda^{\prime \prime},\lambda^{\prime \prime \prime}) \; \psi(\lambda) \psi(\lambda^\prime) \psi(\lambda^{\prime \prime}) \psi(\lambda^{\prime \prime \prime}),
\end{multline}
where
\begin{multline}
A(\lambda,\lambda^\prime,\lambda^{\prime \prime},\lambda^{\prime \prime \prime}) = \\
\int dx_1 dx_2 dx_2^\prime dx_2^{\prime \prime} dx_2^{\prime \prime \prime} \dots dx_{n+1} dx_{n+1}^\prime dx_{n+1}^{\prime \prime} dx_{n+1}^{\prime \prime \prime} \\
\psi_{\lambda_2}(x_3-x_2) \: \psi_{\lambda_2^\prime}(x_3^\prime-x_2^\prime)  \: \psi_{\lambda_2^{\prime \prime}}(x_3^{\prime \prime}-x_2^{\prime \prime}) \\
\: \psi_{\lambda_2^{\prime \prime \prime}}(x_3^{\prime \prime \prime}-x_2^{\prime \prime \prime}) \cdots \psi_{\lambda_n}(x_{n+1}-x_n) \: \psi_{\lambda_n^\prime}(x_{n+1}^\prime-x_n^\prime)  \\
\: \psi_{\lambda_n^{\prime \prime}}(x_{n+1}^{\prime \prime}-x_n^{\prime \prime}) \:  \psi_{\lambda_n^{\prime \prime \prime}}(x_{n+1}^{\prime \prime \prime}-x_n^{\prime \prime \prime}) \; \psi_{\lambda_1}(x_2-x_1) \\
\psi_{\lambda_1^\prime}(x_2^\prime-x_1) \: \psi_{\lambda_1^{\prime \prime}}(x_2^{\prime \prime}-x_1) \: \psi_{\lambda_1^{\prime \prime \prime}}(x_2^{\prime \prime \prime}-x_1) \\
\phi_J(x_{n+1}) \: \phi_J(x_{n+1}^\prime) \: \phi_J(x_{n+1}^{\prime \prime}) \: \phi_J(x_{n+1}^{\prime \prime \prime})
\end{multline}
is the fourth rank constant coupling tensor.  It is only a function of the mother wavelet, $\psi_0 (x)$, and can easily be calculated given that mother wavelet.  Taking the first through fifth functional derivatives gives
\begin{multline}
\frac{\delta S_0[\psi(\lambda)]}{\delta \psi(\lambda)} = (\lambda^2 - m^2) \psi(\lambda) - \frac{\beta}{3!} \int d\lambda^\prime \: d\lambda^{\prime \prime} \: d\lambda^{\prime \prime \prime} \\
A(\lambda,\lambda^\prime,\lambda^{\prime \prime},\lambda^{\prime \prime \prime}) \; \psi(\lambda^\prime) \psi(\lambda^{\prime \prime}) \psi(\lambda^{\prime \prime \prime}),
\end{multline}
\begin{multline}
\frac{\delta^2 S_0[\psi(\lambda)]}{\delta \psi(\lambda) \delta \psi(\lambda^\prime)} = (\lambda^2 - m^2) \: \delta(\lambda - \lambda^\prime) - \frac{\beta}{2!} \int  d\lambda^{\prime \prime} \: d\lambda^{\prime \prime \prime} \\
A(\lambda,\lambda^\prime,\lambda^{\prime \prime},\lambda^{\prime \prime \prime}) \; \psi(\lambda^{\prime \prime}) \psi(\lambda^{\prime \prime \prime}),
\end{multline}
\begin{multline}
\frac{\delta^3 S_0[\psi(\lambda)]}{\delta \psi(\lambda) \delta \psi(\lambda^\prime) \delta \psi(\lambda^{\prime \prime})} = - \beta \int  d\lambda^{\prime \prime \prime} \\
A(\lambda,\lambda^\prime,\lambda^{\prime \prime},\lambda^{\prime \prime \prime}) \;  \psi(\lambda^{\prime \prime \prime}),
\end{multline}
\begin{equation}
\frac{\delta^4 S_0[\psi(\lambda)]}{\delta \psi(\lambda) \delta \psi(\lambda^\prime) \delta \psi(\lambda^{\prime \prime}) \delta \psi(\lambda^{\prime \prime \prime})} = 
- \beta \: A(\lambda,\lambda^\prime,\lambda^{\prime \prime},\lambda^{\prime \prime \prime}),
\end{equation}
and
\begin{equation}
\frac{\delta^5 S_0[\psi(\lambda)]}{\delta \psi(\lambda) \delta \psi(\lambda^\prime) \delta \psi(\lambda^{\prime \prime}) \delta \psi(\lambda^{\prime \prime \prime}) \delta \psi(\lambda^{\prime \prime \prime \prime})} = 0.
\end{equation}

\begin{figure}
\noindent\includegraphics[width=10pc]{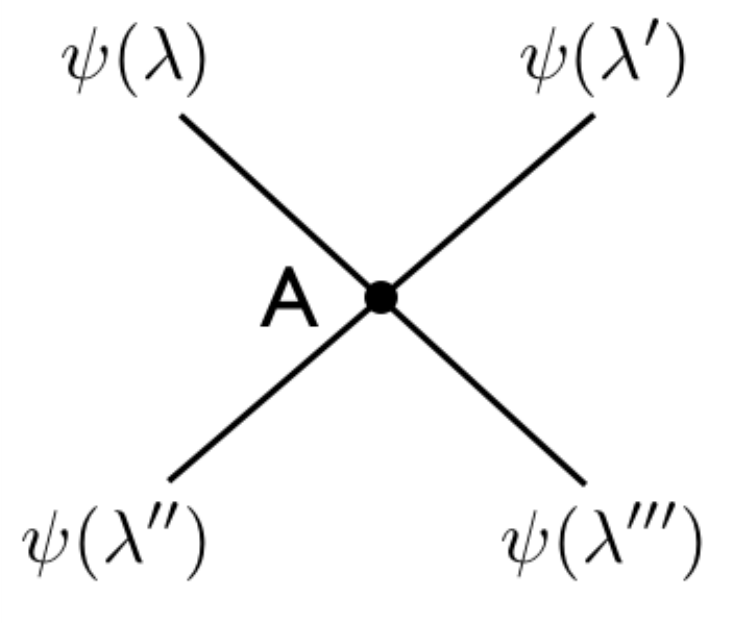}
\caption{\label{fig2} Four point scattering vertex of the $\phi^4$ theory, where $A(\lambda,\lambda^\prime,\lambda^{\prime \prime},\lambda^{\prime \prime \prime})$ is the four point vertex scattering strength.}
\end{figure}

Notice the first term of $S_0[\psi(\lambda)]$ in Eq. \eqref{phi.4.action.eqn} is diagonal in the scattering basis.  This would not be the case if the wavelet basis was not orthogonal.  If the wavelet basis was not orthogonal, it would complicate the first term with off diagonal elements, but it would not prevent further analysis.  The second term explicitly shows the four point vertex scattering (see Fig. \ref{fig2}) with the scattering strength given by $A$.
Note that $A$ is a functional only of the mother wavelet and can be explicitly calculated.  Also note that the fifth order and higher functional derivatives are zero.  This will mean that the fifth order and higher components of the S-matrix must be zero.  Continuing the analysis, expressions for $\varphi_0(\lambda)$ and $S_2^{-1}(\lambda,\lambda^\prime)$ can be derived.  Because this theory only has three scaling functions, $m(\lambda)$, $\beta(\lambda)$, and $f_0(\lambda)$, the dimension of the S-matrix will be limited to $3 \times \dim (\mathbb{R})$ and not have terms greater than fourth order.

The reason for the limited number of non-zero scattering coefficients is directly related to the simple form of the physical actions.  They tend to have a small number of fields, masses and coupling constants ($N_c$);  and the interaction is of low order ($N_i$).  The order of the interaction, $N_i$, is directly related to the maximum order of the scattering operator, and $N_c$ is directly related to the number of scattering coefficients that are independent,
\begin{equation*}
N_c \times \dim (\mathbb{R}) \ll \dim (\mathbb{R}^2),
\end{equation*}
which is a small part of the full space that is available.  This compact support in the scattering basis has the practical advantage that only a small amount of the full scattering operator needs to be calculated and analyzed to understand and identify the physics.

\section{Conclusions}
\label{conclusions}

The meaning of Mallat's scattering operator in the context of a physical system has been explored.  The first major constraint that physics puts on the scattering operator is the composition property that demands that $\hat{f}(x_{n+1}) = e^{(i/\hbar) \hat{H} (x_{n+1}-x_n)} \hat{f}(x_n) e^{-(i/\hbar) \hat{H} (x_{n+1}-x_n)}$.  This allows the scattering operator to be written as a functional derivative of a generating function that is of the form
\begin{equation*}
Z[J] = N \int [d\psi (\lambda)] \; e^{(i/\hbar) S_0[\psi (\lambda)] + (i/\hbar) \int dx J(\lambda) \psi (\lambda)}
\end{equation*}
where $S_0[\psi(\lambda)]$ is the action functional.  This is a nontrivial constraint on the scattering operator.  Since the action is normally given as $S_0[f(x)]$, the important calculation is determining $Z[J(\lambda)]$ given $S_0[f(x)]$.  This was shown to be a successive averaging of the physics (the generalized force and the change in the generalized momentum) over scale.  To leading order, including the quantum fluctuation term, the generating function can be written as a functional of $\varphi_0(\lambda)$ and $S_2^{-1}(\lambda,\lambda^\prime)$ which are the expected values of the first and second order scattering operators, or equivalently the classical action (with quantum fluctuation corrections) as a function of the renormalization scale and a transfer matrix (scale dependent mass of the excitations) as a function of the initial and final renormalization scales. The physics is encoded in these functions.  The correspondence is completed by the identification of the distribution of functions, $F(f)$, as the state of the physical system; the function, $f(x)$, as the field; the $1/J$ cutoff of the wavelet transform as the $i\varepsilon$ convergence factor of the path integral, the necessity and freedom in setting the phase to a constant as the existence of a gauge boson corresponding to exchange of this phase; and, most importantly, the expected value of the scattering operator as the S-matrix.

What $S_0[\psi(\lambda)]$ really expresses is how the physics changes as a function of the scale in inverse distance, $\lambda$.  The simultaneous change of basis from $\hat{f}(x) \to \hat{\psi}(\lambda)$ and $\hat{x} \to \hat{\lambda}$ puts one in the natural basis for calculation of the S-matrix and therefore does not have any singularities to renormalize.

The practical significance of this transformation is the compact representation of the physics, to leading order, in the $\varphi_0(\lambda)$ and $S_2^{-1}(\lambda,\lambda^\prime)$ functions.  This compactness is further enhanced by the relatively few parameters of the physical theory that become functions of scale, and by the low order terms that tend to appear in practical $S_0[f(x)]$.

There are two major classes of application of this perspective on renormalization.  The first is, given an action $S_0[f(x)]$, calculating the S-matrix.  This has great potential to get renormalized solutions of quantum gravity and to allow calculation of QCD in the strong coupling limit.  The second is, in the analysis of the state of the system, to understand the underlying physics.  It is even more practical than this.  Given textures (that is, states) that have come from self organization of complex physical systems, it can be determined which of those textures were generated by the same system, that is by the same physics.  As Mallat has discussed, the scattering transformation (to leading order, $\varphi_0(\lambda)$ and $S_2^{-1}(\lambda,\lambda^\prime)$) gives a natural metric on the space of complex systems.  In other words, given three textures, it can be determined which two are most alike.  A related application is the extrapolation of scale behavior.   If it is known that textures belong to just a few possible classes, those textures could be identified from a very small part of the S-matrix, then the texture could be extrapolated in scale by using the known full S-matrix for that class.

Returning back to the quote from Ken Wilson, he was quite right to have had the instinct to have looked for such a basis.  What he did not have at his finger tips was this scattering operator which combines the best attributes of the Fourier and wavelet transforms, and eliminates the unwanted attributes of both. (This would be a lack of stationarity of the wavelet transform and a lack of Lipschitz continuity to deformation of the Fourier transform manifested as the UV divergence, that is singularity, in the conventional field theory calculation of the S-matrix.)  The scattering transform accomplishes this by an iteration of the wavelet transform.  This is the innovation of this perspective on renormalization.


%

\begin{acknowledgments}

Thanks is given to CSIRO for supporting this research through their Science Leaders Program, and the Institut des Hautes Etudes Scientifique (IHES) for hosting a stay where many of the details of this theory were formalized and the first draft of this manuscript was written.  Thanks is also given to both Stephane Mallat and Joan Bruna for many useful discussions, and communication of many of their mathematical results before they were presented or published.  Finally, thanks is given to the University of Western Australia, John Hedditch, and Ian MacArthur for their help in understanding many of the finer points of Quantum Field Theory.

\end{acknowledgments}

%

%

\end{document}